\documentclass[prl,twocolumn,showpacs,nobibnotes,floatfix,superscriptaddress]{revtex4}

\usepackage{graphicx,eucal}

\begin{document}


\title{Quantum Fidelity Decay of Quasi-Integrable Systems}
\author{Yaakov S. Weinstein}
\thanks{To whom correspondence should be addressed}
\email{weinstei@dave.nrl.navy.mil}
\author{C. Stephen Hellberg}
\email{hellberg@dave.nrl.navy.mil}
\affiliation{Center for Computational Materials Science, Naval Research Laboratory, Washington, DC 20375 \bigskip}

\begin{abstract}
We show, via numerical simulations, that the fidelity decay behavior of 
quasi-integrable systems is strongly dependent on the location of the 
initial coherent state with respect to the underlying classical phase space. 
In parallel to classical fidelity, the quantum fidelity generally exhibits 
Gaussian decay when the perturbation affects the frequency of periodic phase 
space orbits and power-law decay when the perturbation changes the shape of 
the orbits. For both behaviors the decay rate also depends on initial state 
location. The spectrum of the initial states in the eigenbasis 
of the system reflects the different fidelity decay behaviors. In addition,
states with initial Gaussian decay exhibit a stage of exponential decay
for strong perturbations. This elicits a surprising phenomenon: 
a strong perturbation can induce a higher fidelity than a weak perturbation 
of the same type. 
\end{abstract}

\pacs{05.45.Pq, 03.65.Yz}
   
\maketitle

Manifestations of chaos and complexity in the quantum realm have been widely
explored in connection with the correspondence principle between classical and 
quantum mechanics \cite{H1}. An example is a system's response to small 
perturbations of its Hamiltonian. Peres \cite{Peres1,Peres2} conjectured 
that this response serves as an indicator of chaos applicable to both the 
classical and quantum realms. That is, in both realms the behavior 
of fidelity between a state evolved under perturbed and unperturbed dynamics 
depends on whether or not the dynamics is chaotic. 

Peres' conjecture found an experimental venue in nuclear magnetic resonance
(NMR) polarization echoes. In these experiments the initial state of the system
is evolved forward under its internal dipolar Hamiltonian and then inverted 
by a sequence of radio-frequency pulses \cite{RF}. The inverted Hamiltonian, 
however, will be not be an exact reversal of the internal Hamiltonian due to 
pulse imperfections and interactions with the environment. These perturbations
reduce the subsequent echo amplitude which is the measure of fidelity. 

The polarization echo as a means of studying dynamical irreversibility was 
applied in Ref.~\cite{LUP}, where it was noted that the echo decay behavior 
as a function of time can be exponential or Gaussian, depending 
on the molecule under investigation. The connection between these results 
and the exponential fidelity decay predicted for systems exhibiting quantum
chaos \cite{Peres2} was made in Ref.~\cite{UPL}.

Encouraged by these experimental investigations, Jalabert and Pastawski 
\cite{Jala} applied semi-classical analysis to the evolution of what they
termed the Loschmidt echo, or fidelity decay. Their analysis showed that 
for chaotic systems, when the perturbation is strong enough such that 
perturbation theory fails, the fidelity decay is comprised of two 
exponentially decaying terms. The first of these terms is dominant for 
small errors and can be described by the Fermi golden rule \cite{J1,C1}.
The second term is dominant for strong errors, independent of perturbation
strength, and decays at a rate given by the analogous classical system's 
Lyapunov exponent.

The identification of a classically chaotic signature in quantum systems has
led to detailed studies of fidelity decay behavior. For quantum systems that 
are analogs of classically chaotic systems, a number of regimes have been 
identified based on perturbation strength. For weak perturbations, such that 
perturbation theory is valid, the fidelity decay is Gaussian 
\cite{Peres1,J1,V1}. For stronger perturbations, in the Fermi golden rule 
regime, the fidelity decays exponentially with a rate determined by the 
perturbation Hamiltonian and perturbation strength \cite{Jala,J1,P1,J2,Jo}. 
In many systems the rate of the exponential increases as the square of the 
perturbation strength \cite{J1} (see \cite{W1} for an exceptional case) until 
saturating at the underlying classical systems' Lyapunov exponent 
\cite{Jala,C1} or at the bandwidth of the Hamiltonian \cite{J1}. The 
crossover between the various regimes \cite{Cerr1,Cerr2,Wang2} and the 
fidelity saturation level \cite{YSW1,W2} have also been explored. 
Quantum fidelity decay simulations have also been carried out in weakly 
chaotic systems \cite{Wang1}, and at the edge of quantum chaos \cite{YSW2}.

Relationships between fidelity decay behavior and other quantum phenomena
are also found in the literature. These include the Fourier transform
relation between fidelity decay and the local density of states \cite{J1,W3}, 
issues of classical-quantum correspondence \cite{Ben1}, reversibility 
\cite{Cohen1}, and decoherence \cite{C2}. We also note that fidelity decay 
studies can be carried out on a quantum computer \cite{Jo,Poulin}, and 
that the fidelity has been experimentally determined for a three-qubit 
quantum baker's map on a NMR quantum information processor \cite{baker}.

Studies of fidelity decay in quantum systems have spurred interest
in the fidelity decay of classical systems \cite{Eck}. For chaotic 
classical systems it has been shown that the asymptotic fidelity decay can
be either exponential or power-law, analogous to the asymptotic decay of 
correlation functions \cite{BCV2}. Faster than Lyapunov exponential decays have
also been identified \cite{VP}.

The fidelity decay behavior of quantum analogs of non-chaotic or 
quasi-integrable classical systems has received less attention \cite{P1,P3,L1} 
then its chaotic counterpart and has been the subject of some controversy 
\cite{P2,J3}. Using semi-classical arguments, Prosen \cite{P1,P2} 
demonstrated the counter-intuitive result that quantum fidelity decay 
of regular, non-chaotic, evolution is Gaussian, faster than the exponential 
decay of chaotic systems. This was challenged by further semi-classical 
arguments \cite{J3} which indicated a power-law decay. A proposed resolution 
\cite{P3} differentiates between individual minimum uncertainty states, which 
generally exhibit a Gaussian decay, and averages over many such states, 
which may be biased by specific states exhibiting power-law fidelity decay 
behavior. 

In this work, we explore what causes a quantum state undergoing regular quantum
evolution to exhibit Gaussian or power-law fidelity decay behavior. We present 
numerical results demonstrating that the behavior depends on the reaction 
of the underlying classical phase space to the applied perturbation. 
Building off classical fidelity decay results \cite{Ben2}, we chart the 
regions of phase space containing states with initial Gaussian or power-law 
decay. Within the two regions we show that the exact rate of the Gaussian or 
power-law decay is also a function of the coherent state position. 
In addition, a connection is presented between fidelity decay behavior and 
the spectrum of the initial state in the eigenbasis of the system. Finally, 
we probe the dependence of the initial decay behavior on perturbation 
strength, Hilbert space dimension, and note that, for strong perturbations, 
there exists a transitional exponential fidelity decay behavior after initial 
Gaussian decay and before fidelity saturation. 

Perturbing classical Hamiltonian evolution can affect phase space orbits 
in two general ways: the perturbation may distort the shape of the orbit 
or change the frequency of the orbit. Benenti, Casati, and Veble (BCV) 
\cite{Ben2} proposed that in the limit of weak perturbations the classical 
fidelity decay behavior is solely determined by the dominant perturbation 
effect on the phase space orbits. If the dominant effect on a specific 
orbit is to change its frequency, initial wave packets centered in the region 
exhibit Gaussian decay (assuming Gaussian wave packets). This is what would 
be expected from the fidelity of two Gaussian wave packets moving in 
antiparallel directions, or at different speeds, along a specific path. 
If, however, the effect of the perturbation is to change the shape of the 
KAM torus, states centered in the region will exhibit power-law 
fidelity decay. BCV note that they expect similar results in the quantum realm.

Here, we provide numerical evidence that the correspondence between the 
perturbation's effect on phase space and fidelity decay behavior extends to 
quantum systems. Specifically, we show that quantum fidelity decay behavior 
depends on whether an initial coherent state is centered on a phase space 
orbit  whose frequency is changed due to the perturbation, in which case 
the decay will be Gaussian, or an orbit whose shape is distorted by the 
perturbation, in which case the decay will be power-law. Fidelity decay 
simulations under quantum kicked rotor evolution support a 
suspicion of Ref.~\cite{Ben2}, that quantum states are more prone to 
Gaussian decay due to the quantization of the phase space tori. 

The quantum fidelity decay of an initial state $|\psi_i\rangle$ is given by
\begin{equation}
F(t) = |\langle\psi_i|U^{-t}U_p^t|\psi_i\rangle|^2
\end{equation}
where $U$ is the unperturbed evolution, $U_p = Ue^{-i\delta V}$ is the 
perturbed evolution, $\delta$ is the perturbation strength, and $V$ is the 
perturbation Hamiltonian. Our numerical work is centered around kicked maps 
with kick strength $k$ determining whether the evolution is chaotic or 
regular. For the perturbed evolution we employ the same map with a slightly 
different kick strength. Thus, the unperturbed operator is $U = U(k)$, and the 
perturbed operator is $U_p = U(k+\delta_k)$, with perturbation strength 
$\delta_k$.

\begin{figure*}
\includegraphics[height=8.7cm, width=12cm]{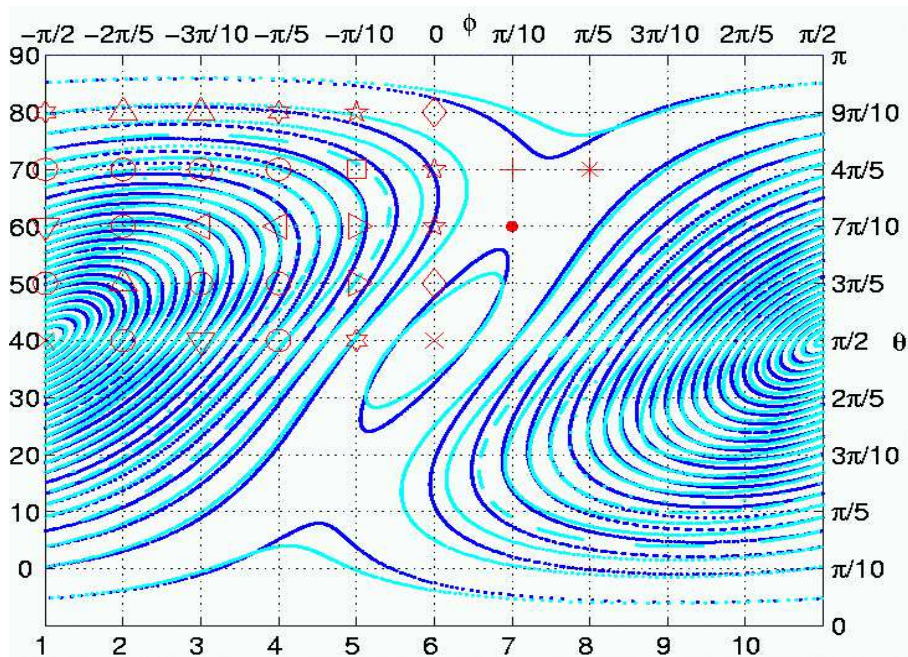}
\caption{\label{CKTphase} 
(Color online) Twenty-five classical orbits on the phase space of the 
classical kicked top, $k_T = 1.1$ (dark), and $k_T = 1.3$ 
(light). The same initial points are used to map the orbits so as to
demonstrate the effect of the $\delta_T$ perturbation. The change of kick
strength primarily affects the frequency of the orbits around the stable 
point at $(\phi = -\pi/2, \theta = \pi/2)$ ($\times$). Thus, under a 
$\delta_T$ perturbation, we expect coherent states placed there to exhibit 
a Gaussian fidelity decay  The change of kick strength does affect the 
shape of the central tori around the $(\phi = 0, \theta = \pi/2)$ ($\times$) 
fixed point. Thus,  under the $\delta_T$ perturbation we expect coherent 
states centered in that region to exhibit a power-law fidelity decay. The 
same holds for the rotational tori at the edges of the phase space. The 
different markers delineate the fidelity decay behavior of coherent states 
centered at those points. The states marked by circles, up, down, left, and 
right pointing triangles, five-pointed stars, and six-pointed stars mark states
that exhibit Gaussian decay behavior at different rates. While we offer no 
clear, a priori, determination of the decay rate, states centered on the same 
orbit tend to decay at similar rates. Though different orbits may also give 
rise to similar decay rates, the general trend is towards a slower Gaussian 
as the states move further from the $(\phi = -\pi/2, \theta = \pi/2)$ fixed 
point. Power-law fidelity decay rates depend on distance from areas with 
large tori distortions, the region surrounding the central fixed point and the 
dip in the rotational torus at the top and bottom of the figure. The closer 
a state is to these regions, the slower the power-law rate. Gaussian and 
power-law fidelity decays are shown in Fig.~\protect\ref{f4} where the 
shapes correspond to the markers used in this figure. 
}
\end{figure*}

We begin our study of fidelity decay with the quantum kicked top (QKT) 
\cite{H2}, a system used in many previous studies of quantum chaos in general 
\cite{H1} and fidelity decay in particular \cite{Peres2,J1,P1,Jo,P3,J3}. The 
classical kicked top describes dynamics on the surface of a sphere
\begin{eqnarray}
x_{t+1} & = &  z_t \nonumber\\
y_{t+1} & = &  x_t \sin(k_T z)+ y_t \cos(k_T z) \nonumber\\
z_{t+1} & = & -x_t\cos(k_T z) + y_t\sin(k_T z),
\end{eqnarray}
where $k_T$ is the kick strength. We choose a kick strength of 
$k_T = 1.1$ corresponding to quasi-integrable dynamics. The $\phi-\theta$
phase space of the classical kicked top is shown in Fig.~\ref{CKTphase} 
where $\theta_t = \arccos(z_t)$ and $\phi_t = \arctan(y_t/x_t)$. The 
phase space has a stable fixed point at $(\phi = -\pi/2, \theta = \pi/2)$ 
surrounded by KAM tori, and rotational KAM tori at the $\theta$-edges. 
Another stable fixed point is found at $(\phi = 0, \theta = \pi/2)$ encircled 
by a smaller region of stable KAM tori.

Fig.~\ref{CKTphase} illustrates the effect of changing the kick strength on
the classical kicked top phase space by plotting orbits of two different 
perturbation strengths. The shapes of the rotational orbits in the regions at 
the $\theta$-edges of phase space and of the tori around the central fixed 
point change significantly while those around the fixed point at 
$(\phi = -\pi/2, \theta = \pi/2)$ do not. If correspondence holds between 
classical and quantum fidelity decay, this observation should alert us as to 
the likely fidelity behavior of coherent quantum states centered in these 
phase space regions. A power-law decay is expected for states centered in 
the former regions, and a Gaussian decay for those centered in the latter 
region.

The quantum kicked top (QKT) \cite{H2} is defined by the Floquet operator
\begin{equation}
U_{QKT} = e^{-i\pi J_y/2}e^{-ik_T J_z^2/2J}.
\end{equation}
where $J$ is the angular momentum of the top and ${\bf \vec{J}}$ are the 
irreducible angular momentum operators. The Hilbert space dimension of 
the top is $N = 2J+1$. The representation is such that 
$J_z$ is diagonal. As initial states we use minimum uncertainty 
angular momentum coherent states centered around $(\phi_i, \theta_i)$ 
\cite{Peres2} and employ a QKT of $J = 500$ unless otherwise noted. 

For convenience we number the states assuming a 10 by 10 grid
evenly spaced in the $\theta$ and $\phi$ directions as seen in Fig.~
\ref{CKTphase}. The lines of the grid are numbered such that the number of 
a state, centered at an intersection of the grid, is determined by adding the 
numerical values of the horizontal (numbers on left) and vertical 
(numbers on the bottom) lines. State 1 is thus located at 
($\phi = -\pi/2$, $\theta = \pi/10$) and state 100 at 
($\phi = 2\pi/5$, $\theta = \pi$). In this way, the fixed point at 
$(\phi = -\pi/2, \theta = \pi/2)$ is number 41 while the fixed point at 
$(\phi = 0, \theta = \pi/2)$ is number 46.

\begin{figure}
\includegraphics[height=5.8cm, width=8cm]{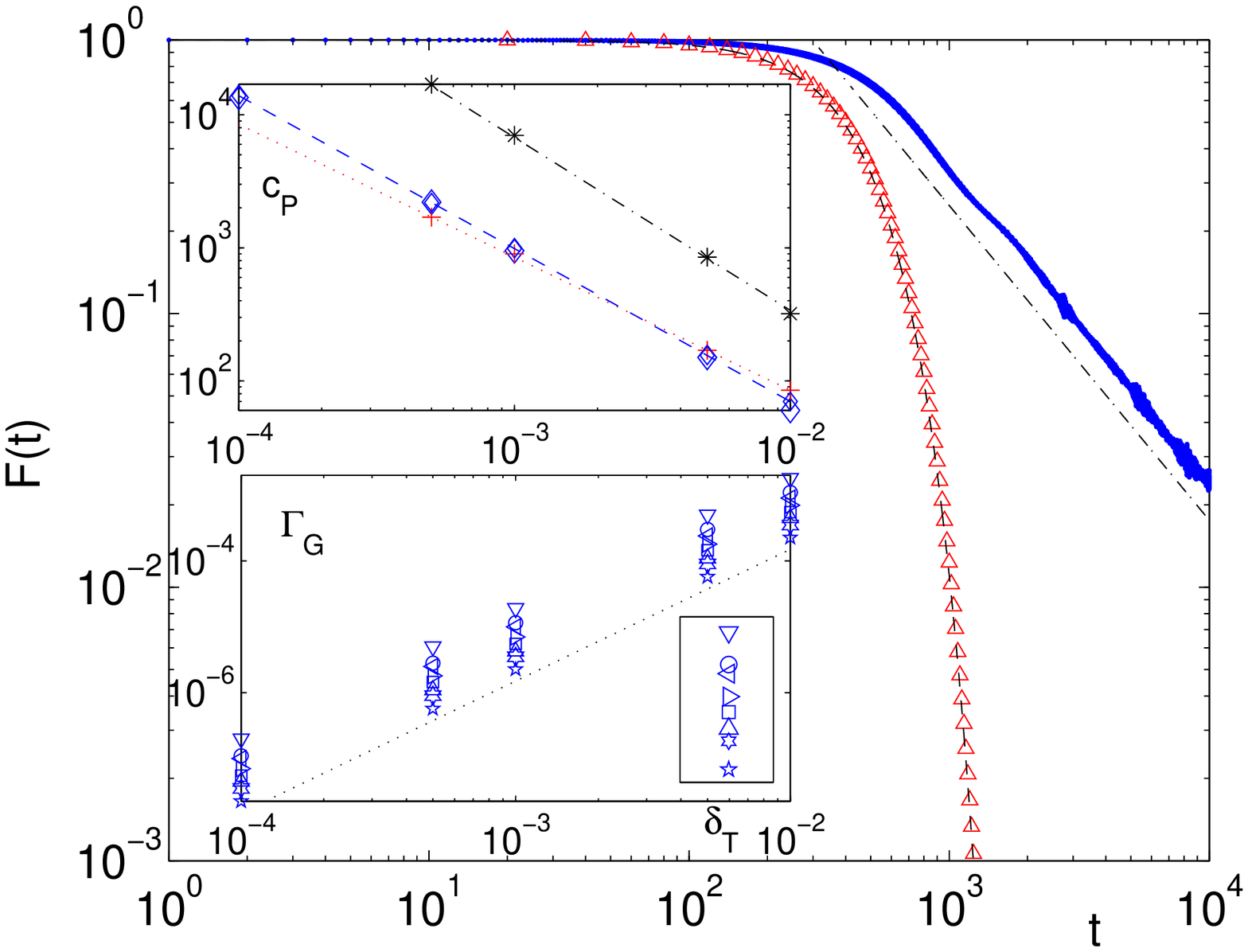}
\caption{\label{f4} 
(Color online) Fidelity decay for quantum kicked top (QKT) with 
$k_T = 1.1, \delta_T = .001$, and $J = 500$, of two different 
coherent states. One (triangles) placed in the region of stable KAM tori 
surrounding the fixed point $(\phi = -\pi/2, \theta = \pi/2)$ (state 52, 
up triangle in Fig.~\protect\ref{CKTphase}, every 20 steps shown) where the 
perturbation effects a change in phase space orbit frequency, and the other 
(solid line) placed in the region surrounding the $(\phi = 0, \theta = \pi/2)$ 
stable point (state 56, diamond in Fig.~\protect\ref{CKTphase}) where the 
perturbation causes a distortion of the phase space orbit. As expected, the 
former displays a Gaussian fidelity decay (dashed line), $e^{-\Gamma_Gt^2}$, 
with $\Gamma_G = 4.5\times 10^{-6}$, while the latter exhibits a power law  
decay $c_Pt^{-\alpha_P}$ (dash-dotted line) with $\alpha_P = 1.15$ and 
$c_P = 950$. The insets show the change of fidelity decay rate as a function 
of perturbation strength, $\delta_T$. The top inset plots $\delta_T$ versus 
$c_P$ for states that decay as $\propto  t^{-1}$ (+), $t^{-1.15}$ (diamond), 
and $t^{-1.3}$ (*) (as explained later, different coherent states decay 
with different power-laws). The points on the loglog plot are well fit by 
$c_P = .85\delta_T^{-1}$ (dotted line), $.35\delta_T^{-1.15}$ (dashed line) 
and $.85\delta_T^{-1.3}$ (dash-dotted line). From these and numerical results 
of other states we assume the following form for power law decay 
$F_P(t) = c_P(\delta_T t)^{-\alpha_P}$. The lower inset shows 
the change of the Gaussian rate, $\Gamma_G$, as a function of $\delta_T$ 
(marked by shapes corresponding to those in Fig.~\protect\ref{CKTphase}). 
As addressed below, coherent states centered in different areas of phase 
space exhibit different Gaussian decay rates, with the rate slowing as the 
states move further from the stable point. The dependence of the rate on 
perturbation strength, however, is always $\Gamma_G \propto \delta_T^2$ 
(dotted line).
}
\end{figure}

Fig.~\ref{f4} shows that state 52, centered in the region of phase 
space surrounding the fixed point $(\phi = -\pi/2, \theta = \pi/2)$, 
exhibits the expected Gaussian behavior (triangles), and that state 66, 
near the fixed point $(\phi = 0, \theta = \pi/2)$ (solid line), exhibits 
the expected power-law decay. These states parallel the expected classical 
fidelity decay behavior.

The existence of states with power-law decay supports the contentions of 
Ref.~\cite{P3} in explaining contradictory results in regular system fidelity 
decay behavior. These states bias the average to look like a power law which 
is slower than the exponential fidelity decay of chaotic systems. Many states, 
however, exhibit Gaussian fidelity decay which may, under certain 
circumstances, be faster than the decay of the corresponding chaotic fidelity. 

\begin{figure}
\includegraphics[height=5.8cm, width=8cm]{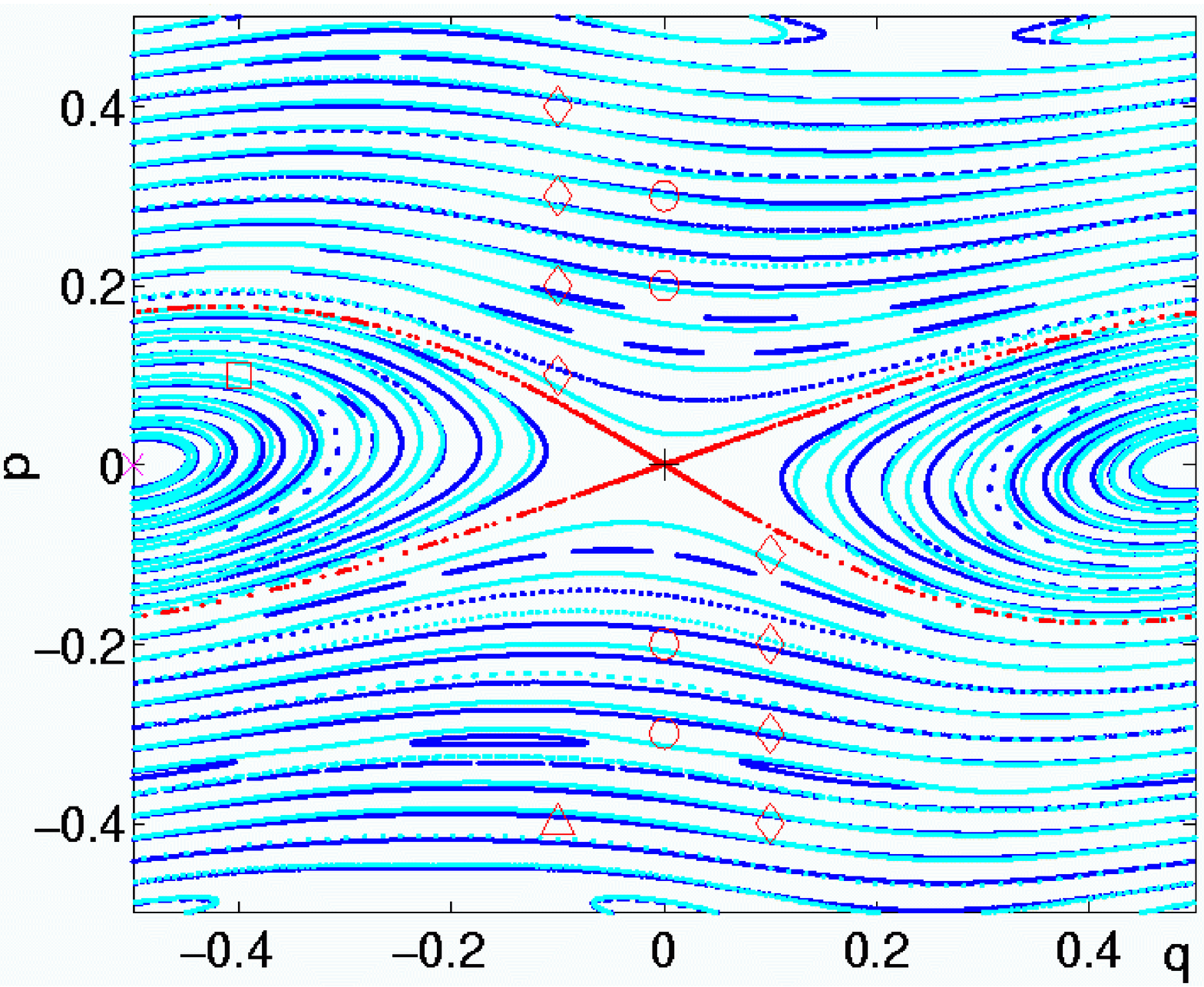}
\caption{\label{CKRphase} 
(Color online) Thirty orbits on the phase space of the classical 
kicked rotors, $k_R = .3$ (dark), and $k_R = .35$ (light). The 
same initial points are used to plot the orbits of both maps in order to 
highlight the effect of a $\delta_R$ perturbation. Well within the 
separatrix, the effect of the $\delta_R$ perturbation is generally to 
change the frequency of the KAM tori. Outside the separatrix, however, 
the $\delta_R$ perturbation essentially changes the shape of the tori. 
Also shown are the stable, $(q = -.5, p = 0)$ ($\times$), and unstable, 
$(q = 0, p = 0)$ (+), fixed points and states used in the text to demonstrate 
Gaussian (square) and non-Gaussian (circle, diamond) fidelity decay. 
Classically, states outside the separatrix are expected to exhibit 
power-law fidelity decay. In the quantum realm, however, we find that 
many such states exhibit Gaussian decay, an example of which is marked 
by the triangle.
}
\end{figure}

The parallel between the fidelity of quantum states and their 
classical counterparts is not, however, the case in general. Some states 
initially centered in areas of apparent phase space orbit distortion 
exhibit the Gaussian fidelity decay expected from perturbations of 
an orbit's frequency. We show this in the quantum version of the system 
explored classically by BCV \cite{Ben2}, the kicked rotor. The classical 
dynamics of the kicked rotor is given by 
\begin{eqnarray}
p_{t+1} & = & p_t+(k_R/2\pi)\sin(2\pi q_t) \nonumber\\
q_{t+1} & = & q_t+p_{t+1}
\end{eqnarray}
where $k_R$ is the rotor kick strength and $-1/2<q,p<1/2$. For $k_R = .3$ 
the phase space of the kicked rotor has a stable fixed point at 
$(q = -.5, p = 0)$ ($\times$) and an unstable fixed point at $(q = 0, p = 0)$ 
(+). The phase space is divided into two distinct regions, 
orbits around the stable fixed point, and rotational motion, as shown in 
Fig.~\ref{CKRphase}. The orbit at the border between these regions is the 
separatrix. As with the QKT we plot phase space orbits of different $k_R$ 
to demonstrate the effect of a change of kick strength 
perturbation on different parts of the classical phase space. The shapes of 
certain KAM tori, such as the ones just outside the separatrix, exhibit 
large deformations while others, such as the inner circles within the 
separatrix, do not. As shown in \cite{Ben2}, states in the former region 
exhibit Gaussian classical fidelity decay while those in the latter region 
exhibit power-law decay. 

To study quantum fidelity decay of the kicked rotor, we use the unitary 
operator describing the quantum kicked rotor (QKR) \cite{L1}
\begin{equation}
U_{QKR} = e^{-ip^2\pi N}e^{-ik_R\cos(2\pi q)N/\pi},
\end{equation}
where $N$ is the Hilbert space dimension. For our simulations we use QKRs of
$N = 500, 1000$, $k_R = .3$, corresponding to a classical kicked rotor with
non-chaotic dynamics, and perturbation strengths $\delta_R = .002, .0014$. 
As initial states we use the minimum uncertainty coherent states described in
\cite{Sar} centered around $(q_i,p_i)$.

\begin{figure}
\begin{center}
\includegraphics[height=5.8cm]{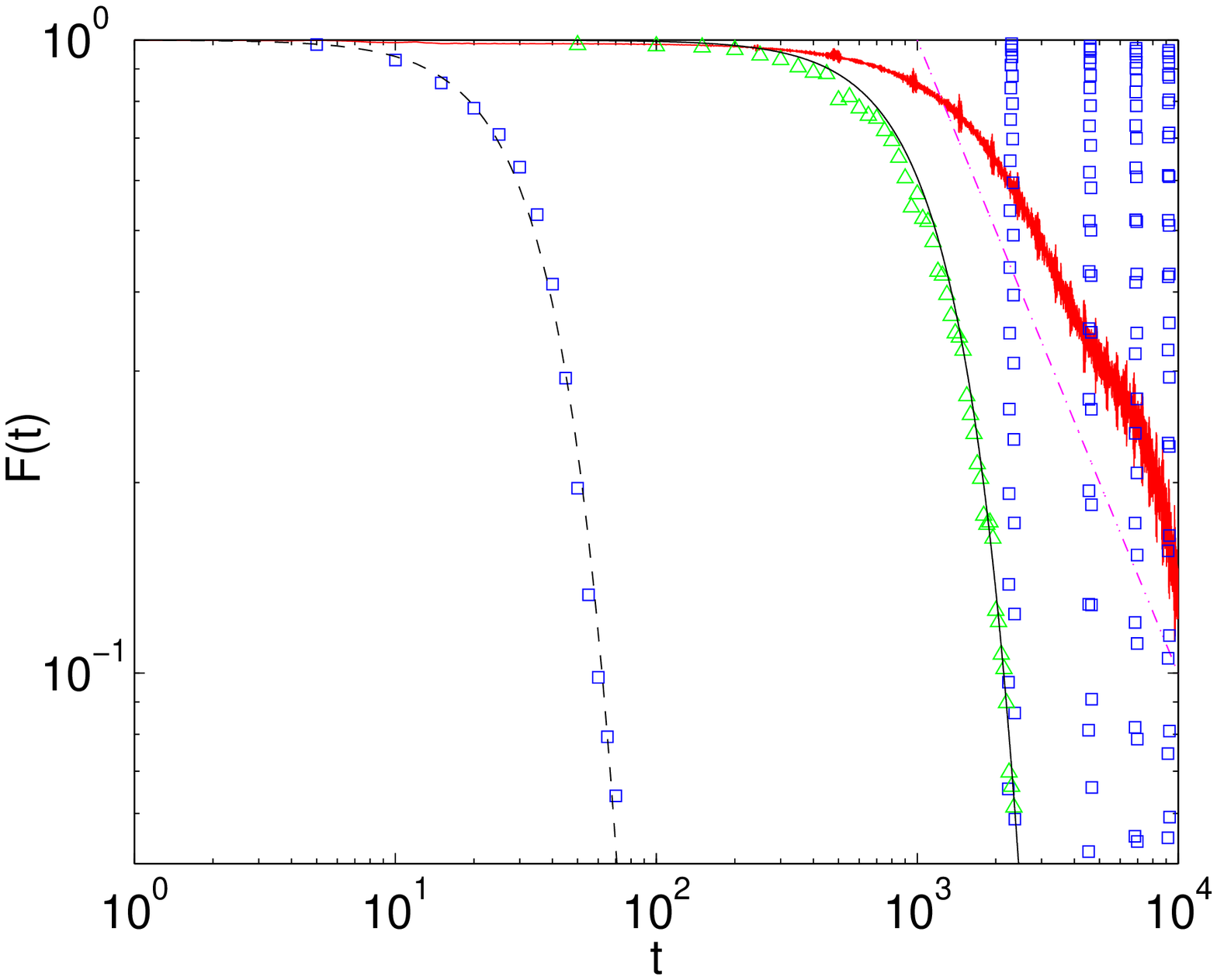}
\includegraphics[height=5.8cm]{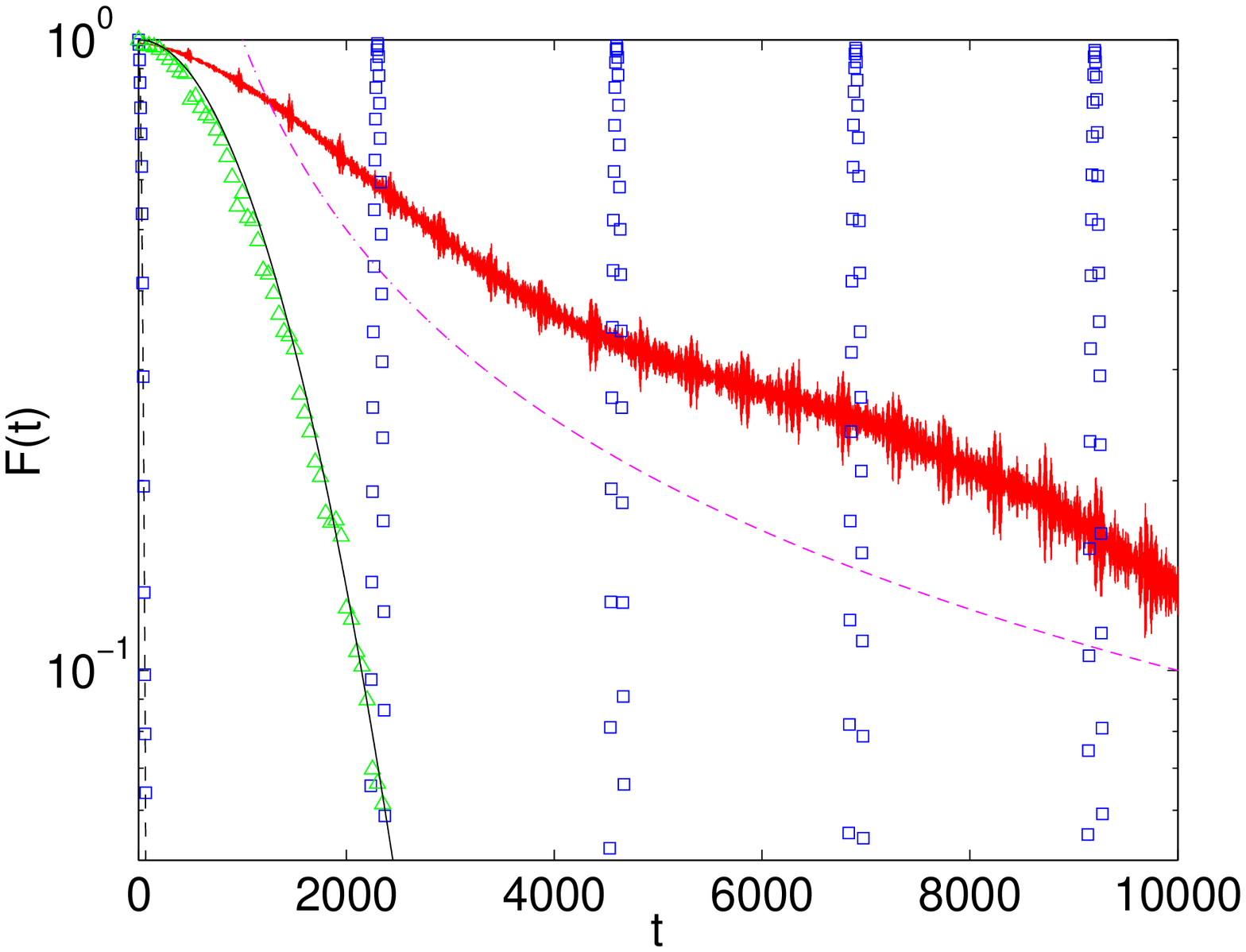}
\end{center}
\caption{\label{f2} 
(Color online) Log-log and log-linear plots of three coherent state fidelity 
decay behaviors under evolution of the quantum kicked rotor with $k_R = .3, 
\delta_R = .002$, and $N = 500$. One of the states is centered inside the 
separatrix at ($q=-.4,p=.1$) (squares every 50 time steps), where there is 
no apparent distortion of KAM tori. The location of this state on the 
classical phase space is marked by a square in Fig.~\protect\ref{CKRphase}. 
Another state is centered outside the separatrix ($q=-.1,p=.1$) (solid line,
diamond in Fig.~\protect\ref{CKRphase}) where there is noticeable KAM tori 
distortion due to the perturbation. The former displays a Gaussian 
fidelity decay (dashed line), $e^{-.0006t^2}$ and exhibits fidelity recurrences
every 2300 time steps \protect\cite{L1}. The latter exhibits a decay
which is non-Gaussian. A power-law  $\propto t^{-1}$ (dash-dotted line) is 
plotted for comparison. The third state (triangle) is centered outside the 
separatrix (triangle in Fig.~\protect\ref{CKRphase}) but nonetheless 
exhibits a Gaussian decay $F(t) = e^{-5\times 10^{-7}t^2}$. This state does
not exhibit fidelity decay recurrences. 
}
\end{figure}

Fig.~\ref{f2} shows the fidelity decay of quantum coherent states 
centered in different regions of phase space. One (squares) in a phase space 
region where the effect of $\delta_R$ is primarily to change the frequency of 
the KAM tori,$(q = -.4, p = .1)$ (square in Fig.~\ref{CKRphase}), and another 
(solid line) in a phase space region where $\delta_R$ primarily distorts the 
shape of the KAM tori $(q = -.1, p = .1)$, (diamond in Fig.~\ref{CKRphase}). 
The former exhibits a Gaussian fidelity decay while the latter exhibits a decay
which is non-Gaussian and resembles (but is not quite) a power-law. 
An exact power-law does not emerge for any of the states simulated for 
the QKR even for perturbation strengths as low as $\delta_R = .0005$. 
Fig.~\ref{f2} also shows a state, $(q = -.1, p = -.4)$, centered on a 
rotational orbit outside the separatrix which, despite the classical 
prediction of a power-law fidelity decay, exhibits Gaussian decay 
(triangle in Fig.~\ref{CKRphase}). 

As mentioned above, BCV \cite{Ben2} predict a lack of correspondence between 
classical and quantum fidelity decay noting that the quantization of KAM 
tori tends to suppress transitions between them. This would enforce a change 
of frequency as the primary effect of the perturbation. The lack of an actual 
power-law decay at any point in the QKR orbits implies such a suppression 
throughout the region of rotational motion. Lack of correspondence to 
classical dynamics for rotational QKR orbits has also been noted with 
respect to fidelity recurrences \cite{L1}. Fidelity recurrences, as seen in 
Fig.~\ref{f2}, are predicted classically in all regions of phase space. Yet, 
in the quantum realm, they do not occur in the QKR rotational tori. 
Ref.~\cite{L1} notes that these regions have a high density of states and 
the enhanced quantum interference may cause the lack of classical 
correspondence. A similar argument can be proposed here, as enhanced 
quantum interference may be especially sensitive to perturbations and 
cause the fast Gaussian fidelity decay.

In attempt to further understand what differentiates states that exhibit 
Gaussian fidelity decay from those that exhibit power-law decay we look at 
the spectra of the initial coherent states with respect to the QKT eigenbasis 
as a function of the extent of the eigenstates in $J_z$. Following Peres 
\cite{Peres2} the extent of a state with respect to the operator $J_z$ is 
defined as
\begin{equation}
\Delta|J_z| = \sqrt{\langle\psi|J_z^2|\psi\rangle-|\langle\psi|J_z|\psi\rangle|^2}.
\end{equation}
The extent is basically the first term in the power series expansion of 
the fidelity \cite{Peres1} and thus we expect it to provide insight into 
the expected fidelity decay behavior. We calculate the extent of all the 
eigenstates of the QKT and see how much each of these states contributes 
to a given coherent state. The contribution is quantified by an amplitude 
$A_j = |\langle\psi_i|\phi_j\rangle|^2$ where $\psi_i$ is the initial 
coherent state and $\phi_j$ is the $j$th QKT eigenstate.

On the extremes, the coherent state centered at the stable fixed point 
$(\phi = -\pi/2, \theta = \pi/2)$ is primarily ($A = .95$) composed of one 
eigenstate with $\Delta|J_z| = 17.3$. The primary contributors
to the coherent state centered at the stable fixed point at
$(\phi = 0, \theta = \pi/2)$ are 4 eigenstates each with amplitudes of .21
and $\Delta|J_z| = 353.7$. In both cases the fidelity barely decays as the 
state lives in a constricted Hilbert space \cite{Peres2,P1,YSW1}.

\begin{figure}
\includegraphics[height=5.8cm, width=8cm]{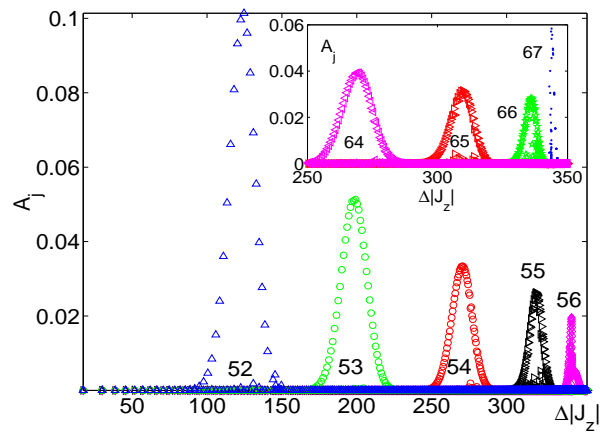}
\caption{\label{Ext} 
(Color online) Contribution of QKT eigenstates for coherent states 52-56 (left 
to right) versus extent of those eigenstates in $J_z$ with shapes as in Fig.~
\protect\ref{CKTphase}. As the states move further from the low extent 
$(\phi = -\pi/2, \theta = \pi/2)$ fixed point more eigenstates have significant
contributions and the height of the Gaussian contribution curve decreases. 
Consequently, the Gaussian fidelity decay rate decreases. State 56, made of 
only high-extent states, does not have a Gaussian extent spectrum and 
exhibits power-law fidelity decay. The inset shows coherent states 64-67 
(left to right) which follow a similar pattern.
}
\end{figure}

\begin{figure}
\includegraphics[height=5.8cm, width=8cm]{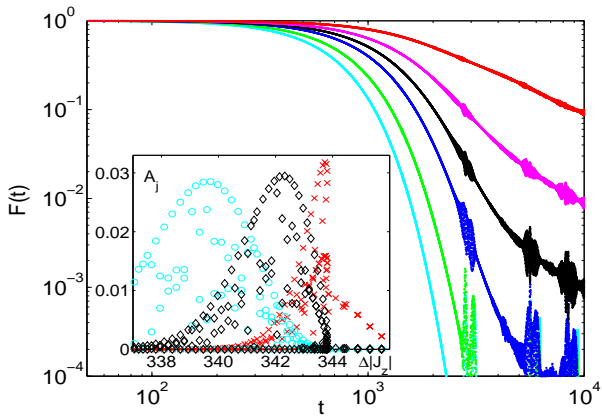}
\caption{\label{G2P} 
(Color online) Fidelity decay of coherent states with $\theta = 4\pi/5$ and 
$\phi$ ranging from $0$ (state 76) and $\pi/10$ (state 77) for the QKT with 
$k_T = 1.1$, $\delta_T = .001$. The states shown are 
for $\phi = \pi/100, 3\pi/100, 5\pi/100, 6\pi/100, 7\pi/100$, and  $9\pi/100$ 
(bottom to top). The decay can be seen to transition smoothly from Gaussian 
to power-law. The inset shows the extent spectrum of the states at 
$\phi = 3\pi/100$ (circles), $6\pi/100$ (diamonds), and 
$9\pi/100$ ($\times$). As the power-law fidelity decay region is 
approached, the Gaussian spectrum gets filled in and narrows on the side of 
higher extent becoming almost flat except for a tail of high-extent states.
}
\end{figure}

Most coherent states, however, have significant contributions from many 
different eigenstates. In Fig.~\ref{Ext} the contribution to coherent states 
52-56 is plotted versus the extent of the eigenstates and in the inset the 
same is plotted for states 64-67. The general pattern emerging from the 
figure (and from states not shown) is clear. Coherent states exhibiting 
Gaussian fidelity decay have a Gaussian spectrum of contributions from 
eigenstates with low to middle range extents in $J_z$. As the coherent 
states move away from the low-extent
$(\phi = -\pi/2, \theta = \pi/2)$ fixed point (with distance determined by 
the number of passing trajectories) more and more states contribute
and with lower amplitudes, higher extent, and a more localized extent range. 
The Gaussian shape remains until the coherent states enter the region 
of power-law fidelity decay. States with power-law fidelity decay have a 
very narrow spectra at high extent. The difference between the types of 
states is clearly seen in Fig.~\ref{Ext} and the transition between the types
of states can be seen in Fig.~\ref{G2P}. 

We suggest that the relationship between fidelity decay and the extent 
spectrum may be understood as follows. States exhibiting power-law decay 
have a large extent in the direction of the perturbation, $J_z$. When the 
coherent state is perturbed these states cannot spread out much more in the 
perturbation direction. Rather, they interfere with each other and the decay 
is slow. Coherent states exhibiting Gaussian decay, however, are spread out 
in extent space. The perturbation affects each of these states differently 
spreading them out in $J_z$ and causing a ballistic decay. As the coherent 
states move away from the low extent fixed point the average extent of 
these states grows and the Gaussian fidelity decay gets slower until the 
transition to power-law decay. This description holds for states in the 
Gaussian and power-law fidelity decay regions for the QKT phase space. 
States at the border between these regions and states very close to the 
fixed points have different extent spectra and, thus, exhibit fidelity 
decay behavior that is neither Gaussian nor power-law. These regions 
will be discussed below.

A full exploration of the extent and its relation to fidelity decay is 
beyond the scope of this paper. However, looking at the spectrum of 
a coherent state as a function of the extent of the contributing basis states 
with respect to the perturbation operator, $V$ (or some function thereof), can 
help identify the regions of different decay behaviors.

We now embark on a more extensive exploration of coherent state fidelity 
decay behavior in the regular regime of the QKT. To this aim, we have 
calculated the fidelity decay for coherent states spaced throughout the 
classical phase space for a number of perturbation strengths and Hilbert space
dimensions. A large variety of behaviors exist, though we 
concentrate only on the initial decay before any fidelity recurrences.
In attempt to organize the data in a straightforward fashion we explore 
fidelity decay as it relates to the following variables: 
perturbation strength, Hilbert space dimension, and position of the initial 
state with respect to the underlying classical phase space. We also study 
hitherto unobserved exponential decay which may occur after an initial 
Gaussian decay and explore how this decay regime behaves with respect to the 
above variables. We note that an extensive semi-classical treatment of 
regular fidelity decay has been done in Ref.~\cite{P1}. Our purpose here is to 
outline an approach based on knowledge of the system's classical phase space. 

We first address the dependence of the fidelity decay rate as a function 
of perturbation strength. For coherent states exhibiting a Gaussian fidelity 
decay, $F_G(t) = e^{-\Gamma_Gt^2}$, numerical simulations verify 
$\Gamma_G \propto\delta_T^2$, as derived in \cite{P1}. This dependence is 
demonstrated in the lower inset of Fig.~\ref{f4} for 
$\delta_T = .0001, .0005, .001, .005, .01$ and $J = 500$. 
For coherent states exhibiting a power law decay, $F_P(t) = c_Pt^{-\alpha_P}$, 
numerical results for the above perturbation strengths suggest that 
$c_P$  is $\propto\delta_T^{-\alpha_P}$, from which we conclude 
$F_P(t) = c(\delta_T t)^{-\alpha_P}$. The upper inset in Fig.~\ref{f4} 
demonstrates this behavior with states whose power-law decay rate is 
$\alpha_P = 1,1.15,$ and $1.3$.

\begin{figure}
\includegraphics[height=5.8cm, width=8cm]{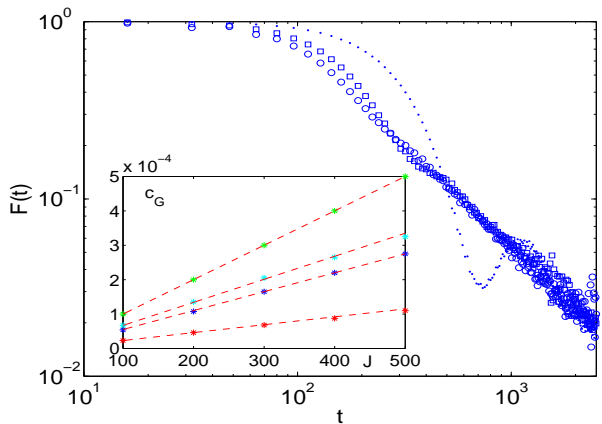}
\caption{\label{JPower} 
(Color online) Coherent state 56 (diamond in Fig.~\protect\ref{CKTphase}) 
power-law fidelity decay (every 16th time step plotted) for the QKT with 
$k_T = 1.1$, $\delta_T = .005$, and $J = 500$ (circles), 300 (squares), 
and 100 (dots). Lowering $J$ causes the fidelity decay behavior to transition 
from power law to Gaussian. The inset shows the rate of Gaussian fidelity 
decay, $\Gamma_G$, as a function of $J$ for some coherent states. 
We find a linear relationship $\Gamma_G = c_G J$ with 
$c_G = 2.3\times 10^{-7}$, $5.5\times 10^{-7}$, $6.7\times 10^{-7}$ 
and $1\times 10^{-6}$ (bottom to top) shown.
}
\end{figure}

To address the fidelity decay behavior as a function of Hilbert space 
dimension we choose one perturbation strength, $\delta_T = .005$, for 
$J = 100, 200, 300, 400, 500$. The fidelity decay is calculated 
for coherent states of appropriate dimension centered at specified points 
in phase space. For states centered in regions of Gaussian fidelity decay 
we find a linear relation between $J$ and $\Gamma_G$, as shown in the inset 
of Fig.~\ref{JPower}, the slope of which depends on the coherent 
state's location in phase space. 

We can thus write the following equation for the Gaussian fidelity decay 
behavior: 
\begin{equation}
F_G(t) = e^{-\Gamma_Gt^2}, \;\;\;\;\;\;\;\; \Gamma_G = \gamma_G J \delta^2
\end{equation}
where $\gamma_G$ depends only on the initial coherent state's location in 
phase space and is the only term not a priori calculable from our analysis. 
This result is in consonance with the semi-classical approach outlined 
in \cite{P1}.

For coherent states in regions of power-law fidelity decay, we find no change 
in decay rate with $J$ as long as the fidelity decay remains a power-law.
Thus we write the following equation for the power-law decay behavior
\begin{equation}
F_P(t) = c(\delta_T t)^{-\alpha_P}
\end{equation}
where $c$ and $\alpha_P$ depend on the coherent state's location.
However, as $J$ is decreased the fidelity decay behavior does change; it 
shifts from power-law to Gaussian, as shown in Fig.~\ref{JPower} (for state 
56, marked diamond in Fig.~\ref{CKTphase}). This shift is due 
to the increasing size of the initial coherent state making it more likely 
that the states will overlap with KAM tori on whom the perturbation effects 
a change of frequency. Coherent states centered more deeply in the power 
law decay region (such as state 77 marked plus in Fig.~\ref{CKTphase}) 
have a slower transition to Gaussian decay when decreasing $J$. 

As we have seen, the fidelity decay behavior in general, and the rate of 
$F_G(t)$ and $F_P(t)$ specifically, are dependent on the exact location 
of the initial coherent state with respect to the underlying classical phase
space. This dependence is emphasized in Fig.~\ref{CKTphase} by using 
different shapes to mark the center of coherent states exhibiting Gaussian 
(circle, square, up, down, left, and right triangles, five-pointed star, 
and six-pointed star) and power-law (diamond, +, dot, *) fidelity decay. 
States with practically equivalent decay rates are represented by the same 
shape, with the rates themselves shown in Fig.~\ref{f4}. Based 
on our simulations we cannot formulate clear cut rules for the decay rate of a 
given coherent state. However, we make two observations. First, states along 
the same phase space orbit tend to have similar decay rates. Second, 
$\Gamma_G$, the rate of Gaussian decay, decreases as the states get further 
from the $(\phi = -\pi/2, \theta = \pi/2)$ fixed point (with distance measured
by the number of trajectories between the fixed point and the center of the 
coherent state). We have already seen consequences of this latter observation 
in the extent spectra. Similarly, for states exhibiting power-law decay 
behavior, the power increases as the states get further from regions of 
large KAM torus distortion.

We now explore the fidelity decay of states at the border between the Gaussian 
and power-law phase-space regions and of states close to the fixed points. 
These regions exhibit a variety of decay behaviors which are reflected in the 
extent spectrum. Looking at the transition from Gaussian to power-law decay
away from the fixed points we note that the transition is a rather smooth 
one in both the decay behavior itself and the extent of the contributing 
eigenstates. Fig.~\ref{G2P} displays these behaviors for coherent states 
$\theta = 4\pi/5$ and $\phi$ ranging from $0$ to $\pi/10$ (between states 76 
and 77 of Fig.~\ref{CKTphase}). For the fidelity, 
the decay slows as the state leaves the region where the 
dominant perturbation effect is on the frequency of the orbits and enters the 
region where the dominant perturbation effect is on the shape of the orbits. 
In the extent spectra the transition is manifest by the Gaussian shape 
narrowing on the side of large extent eventually becoming almost flat 
except for a tail reaching towards the higher extent eigenstates. Similar 
behavior is found for other states in the border region between Gaussian and 
power-law fidelity decay.

\begin{figure}
\includegraphics[height=5.8cm, width=8cm]{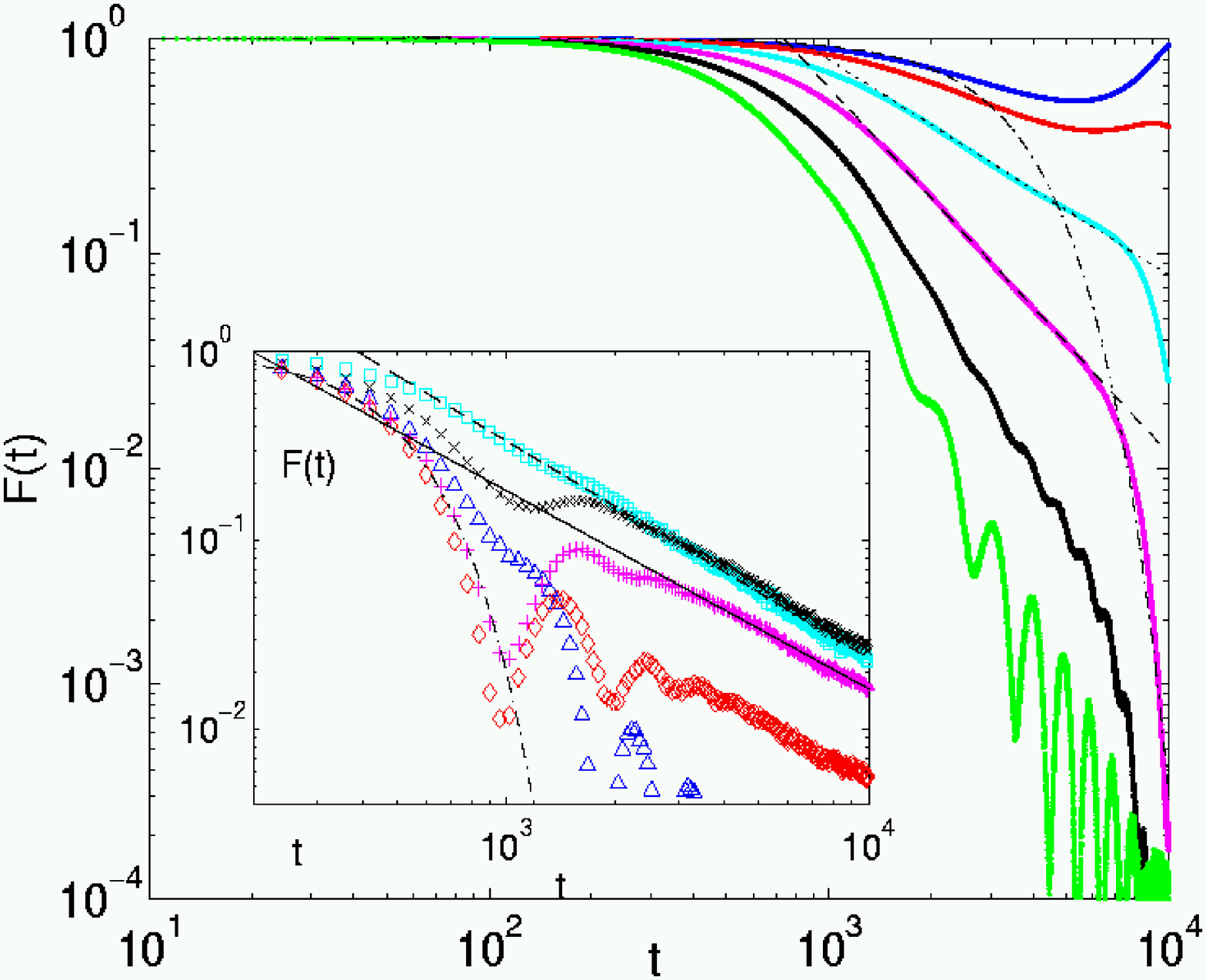}
\includegraphics[height=5.8cm, width=8cm]{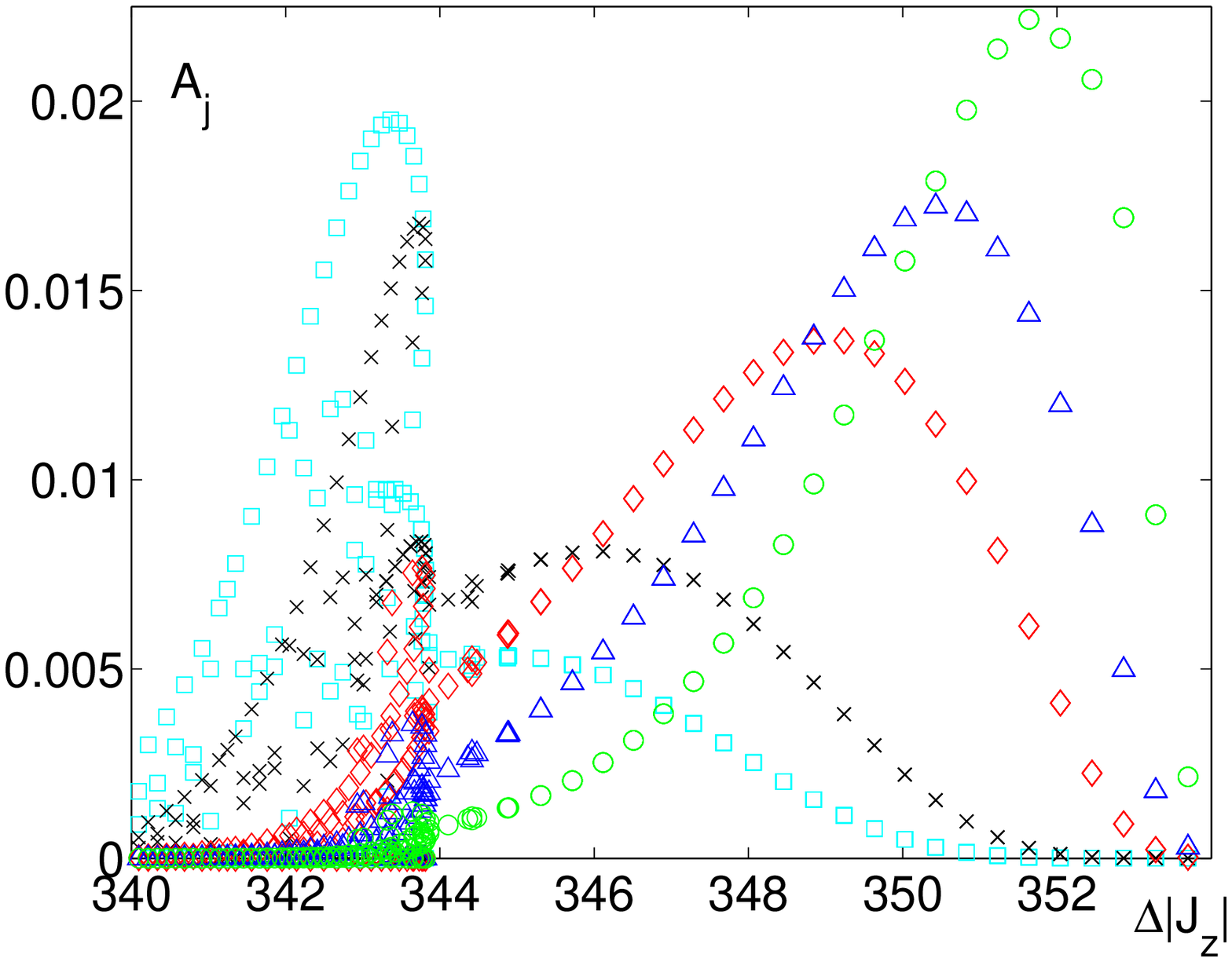}
\caption{\label{C45} 
(Color online) The top panel shows the fidelity decay of coherent states 
with $\theta$ ranging from $\pi/2$ to $6\pi/10$ and $\phi = 0$ for the QKT 
with $k_T = 1.1$, $\delta_T = .001$. The states shown are for 
$\theta = \pi/2 + 0, \pi/100, 2\pi/100, 3\pi/100, 4\pi/100$, and $5\pi/100$, 
(top to bottom). Close to the fixed point the fidelity oscillates
close to one. As the states move further away from the fixed point, 
the fidelity decays in two stages, starting as a power-law and becoming 
a Gaussian. For comparison we fit the initial decay of the 
$\theta = \pi/2 +  2\pi/100$ state with a power-law $\propto t^{-1}$ 
(dotted line), the $\theta = \pi/2 +  3\pi/100$ state 
initial decay with a power-law $\propto t^{-1.7}$ (dashed line), and the 
second stage decay with a Gaussian $e^{-\Gamma_{ss}t^2}$ with 
$\Gamma_{ss} = 8\times 10^{-8}$ (chained line). Moving further away, 
the initial power-law becomes more Gaussian while the second stage starts 
flattening to power-law. The inset shows every 60 steps of the fidelity 
decay for the states $\phi = 0$, $\theta = \pi/2 + 6\pi/10$ (triangles), 
$7\pi/100$ (diamonds), $8\pi/100$ (+), $9\pi/100$ ($\times$), and  
$\pi/10$ (state 56, squares). Here, the initial decay starts off as 
Gaussian and rebounds into a power-law decay. For comparison we plot the 
Gaussian $e^{-\Gamma_{G}t^2}$ with $\Gamma_G = 3.9\times 10^{-6}$ 
(chained line) and the power-law $\propto t^{-1.05}$ (solid line) . 
As the states continue to move away from the fixed point, however, the 
Gaussian flattens until there is a single-behavior power-law decay, 
$\propto t^{-1.15}$ (dashed line). The lower panel shows the extent 
spectrum of the states with $\theta = \pi/2 + 5\pi/10$ (o), $6\pi/10$ 
(triangles), $7\pi/10$ (diamonds), $9\pi/10$ ($\times$), and state 56 
(squares). As the coherent states are moved further away from the fixed 
point the spectra flatten at higher extent eigenstates and a bulge grows 
at lower extent eigenstates.
}
\end{figure}

The region surrounding the  $(\phi = 0, \theta = \pi/2)$ fixed point contains
states exhibiting fidelity decay behaviors not described by a Gaussian or 
power-law and extent spectra different from those seen above. 
Starting with the coherent state centered at the fixed point, the fidelity 
oscillates close to one as the coherent state is comprised almost entirely 
of the highest extent QKT eigenstates. As the coherent states move away from 
the fixed point the highest extent eigenstates still gives the largest 
contributions while the next highest extent eigenstates give increased 
contributions. The fidelity decay in these regions starts off as a power-law 
but exhibits a second stage of Gaussian decay similar to edge of quantum
chaos decays \cite{YSW2}. Moving further, eigenstates with lower and lower 
extent become dominant. However, the extent spectrum is not Gaussian, as 
would be expected for states exhibiting Gaussian fidelity decay, but is 
extended on the side of lower extent eigenstates. At this 
increased distance from the fixed point, the first-stage fidelity decay 
transitions from power-law to Gaussian, and the second-stage decay 
transitions from Gaussian to power-law decay. Finally, the initial Gaussian
fidelity decay flattens into one stage of power-law decay while the extent 
spectrum continues flattening in the direction of higher extent states while 
forming a complicated flattened bulge at lower extent states. All of these 
behaviors are exhibited in Fig.~\ref{C45}. Similar behavior is found in 
regions of coherent states 3-5 and 87-89. 

Coherent states in the region surrounding the fixed point at 
$(\phi = -\pi/2, \theta = \pi/2)$ exhibit behavior that is slightly
different from the states in the region surrounding the 
$(\phi = 0, \theta = \pi/2)$ fixed state. At the fixed point the fidelity 
simply oscillates close to 1. As the coherent states move
away from the fixed point the oscillations become larger in amplitude, the 
recurrence time increases, the initial decay becomes more Gaussian, and the 
maximum fidelity reached on the recurrence is lower. This continues until 
full Gaussian decay behavior emerges. The extent spectra reflect this 
behavior, going from a dominant low extent state to an eventual Gaussian 
shape. 

\begin{figure}
\includegraphics[height=5.8cm, width=8cm]{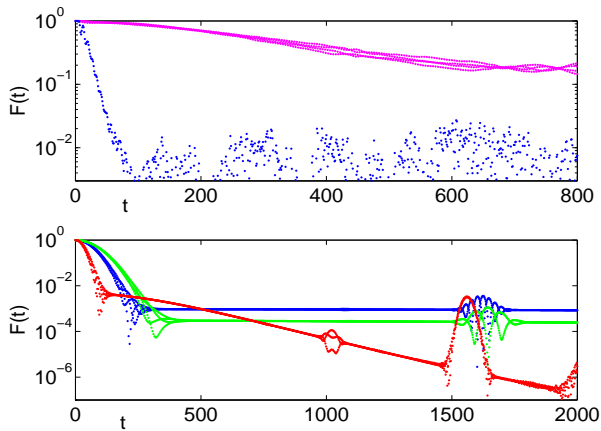}
\caption{\label{Others} 
(Color online) The upper plot shows two coherent states evolved under the 
$N = 1000$, $k_R = .3$ QKR with $\delta_R = .0014$ that exhibit initial 
exponential decay despite the fact that the QKR is in the regular regime. 
The lower plot shows three coherent states under the same QKR evolution which 
exhibit initial Gaussian decay that transitions to exponential decay. For two 
of the states the fidelity freezes after the initial Gaussian decay. The 
fidelity of all three states exhibit echo resonances similar to those 
discussed in Ref.~\protect\cite{P3} for ergodic perturbations.
}
\end{figure}

We also note the presence of states with unexpected fidelity decay behavior
in the QKR. Fig.~\ref{Others} shows examples of $N = 1000$ coherent states 
under QKR $k_R = .3$, $\delta_R = .0014$ evolution that exhibit initial 
exponential fidelity decay (top plot), though the QKR is regular, and fidelity 
`freeze' as discussed in Ref.~\cite{P3}, though the perturbation is 
non-ergodic. 

Beyond the initial Gaussian fidelity decay of some coherent states, there
may exist a second, slower, stage of exponential fidelity decay behavior, 
$F(t) =  c_Ee^{-\beta_E}$, before saturation. This stage is prevalent 
for strong perturbations but disappears for smaller perturbations (or 
smaller Hilbert space dimension with the same perturbation strength). 
The specifics of this exponential decay depend strongly on the phase space 
location of the initial coherent state. The slower exponential decay of this 
second stage gives rise to an exciting phenomenon: a stronger perturbation 
leading to a higher fidelity than a weaker perturbation of the same type. 

The golden rule exponential fidelity decay term mentioned above for 
chaotic systems exists also in regular systems \cite{J3}. We do not identify 
this term with the exponential observed here since, as we show, the exponential
here is strongly dependent on the initial state.

As with the initial fidelity decay behavior we attempt a systematic 
numerical analysis of the second stage exponential decay. We first study 
the exponential as a function of perturbation and then explore the effect
of the location of the initial coherent state. 

\begin{figure}
\includegraphics[height=5.8cm, width=8cm]{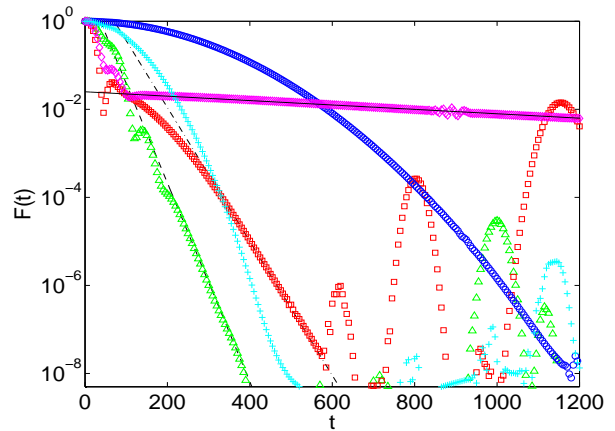}
\caption{\label{G2Exp} 
(Color online) Coherent state 54 (circle in Fig.~\protect\ref{CKTphase})
fidelity decay (every 5th time step plotted) for the QKT with 
$k_T = 1.1$, and $J = 500$, for perturbation strengths $\delta_T = .01$ 
(squares), $.0075$ (diamonds), $.005$ (triangles), $.0025$ (+), and 
$.001$ (o). The fidelity decay of the stronger perturbations show a 
stage of exponential decay after the initial Gaussian. This stage is fitted
by $F(t) = 12e^{-.035t}$ (dash-dotted line) for $\delta_T = .01$, 
$F(t) = .025e^{-.00115t}$ (solid line) for $\delta_T = .0075$, and  
$F(t) = 7e^{-.052t}$ (dashed line) for $\delta_T = .005$. For the strongest 
perturbations, $\delta_T = .01, .0075$, there is also a period of transfer 
between the two fidelity decay regimes, while for $\delta = .005$ this 
transition period is non-existent. For weaker perturbations the exponential 
stage of fidelity decay disappears altogether. The faster Gaussian decay of
the weaker perturbations and the wide range of exponential decay rates, 
leads to the counter-intuitive result that a stronger perturbation leads to 
a higher fidelity as seen in the region $200 < t < 600$ for the 
$\delta_T = .01$ decay and across the whole plotted region for the 
$\delta_T = .0075$ decay. There is no clear correlation between perturbation 
strength and exponential decay rate.
}
\end{figure}

Fig.~\ref{G2Exp} demonstrates that weaker perturbations (+,o) exhibit no 
exponential fidelity decay stage. Rather, the Gaussian 
decay continues until fidelity saturation. As the perturbation strengthens
the second-stage exponential emerges. In addition, there exists a transition 
period between the two decay behaviors.  Thus, weaker perturbations 
lead to longer times of Gaussian fidelity decay during which the fidelity 
of stronger perturbations may have already transferred to the 
slower exponential. In this way, there may be a significant amount of time 
in which the fidelity of the stronger perturbation (diamonds, triangles) is
actually \emph{higher} than that of the weaker perturbation (squares, 
+). This phenomenon is shown in Fig.~\ref{G2Exp}  for the 
$k_T = 1.1$, $J = 500$, QKT. An exponential region of decay is manifest for 
perturbation strengths $\delta_T = .01, .0075,$ and $.005$, but not for weaker 
perturbations $\delta_T = .0025$ and $.001$. Thus, for times $t > 300$ the 
fidelity of at least one of the stronger perturbations is higher than 
the fidelity of a weaker perturbation. 

The rate of the exponential also depends on the perturbation strength, 
$\delta_T$. However, our numerical simulations do not show any simple 
relationship between the exponential rate and the perturbation strength. 
Rather, the rate changes drastically ranging from practically zero, decay 
freeze, to a fast exponential. This also allows a stronger perturbation
to have a higher fidelity than a weaker one. This is exemplified in 
Fig.~\ref{G2Exp} by the $\delta = .0075$ perturbation whose fidelity 
decays very slowly and, thus, after a time, is higher than the fidelity of 
all of the weaker perturbations.

The possibility of a stronger perturbation leading to a higher fidelity 
may have important consequences for quantum simulations in which a quantum 
system is trying to simulate a given dynamics: a strong error in the dynamics 
may be easier to correct via quantum error correction than a weak one. 
This could allow for an interesting error-correction scenario. A weak error 
strongly affecting a system should be purposely \emph{strengthened} so as to
more accurately perform the desired simulation. This would be especially 
significant in a case where the effect of the error is too strong for 
conventional quantum error techniques but can be brought below the 
error-correction threshold if the error is strengthened.

The existence of the exponential fidelity decay region may be related to the 
quantum freeze of fidelity discussed in \cite{P3} for 
ergodic perturbations. In fact, the lower plot of Fig.~\ref{Others} 
displays the fidelity decay of coherent states evolved by the QKR which 
actually freeze, though the applied perturbation is non-ergodic.

\begin{figure}
\includegraphics[height=5.8cm, width=8cm]{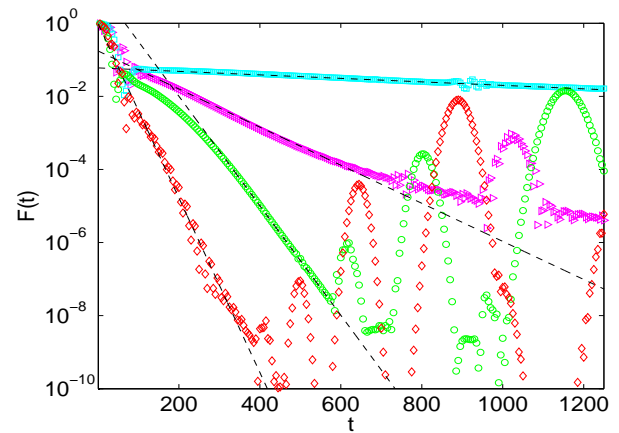}
\caption{\label{G2Exp2} 
(Color online) Fidelity decay for different coherent states (every 5th 
time step plotted) for the quantum kicked top with $k_T = 1.1$, 
$J = 500$, and $\delta_T = .01$. After the initial Gaussian the fidelity 
decay transitions to an exponential, the rate of which depends strongly on the 
location of the coherent state with respect to the underlying classical 
phase space. The states shown are 53 (diamonds), 54 (o), 55 (triangles), 
and 74 (squares). All of these states have initial Gaussian fidelity decay, 
as seen in Fig.~\protect\ref{f4}, which transitions into an exponential 
decay. The current figure exhibits the wide range of exponential decay rates 
and transition times that can occur. Starting with the lowest plot we find
exponential decays of $F(t) = e^{-.055t}$ (diamonds), $10e^{-.0345t}$ 
(circles), $.175e^{-.012t}$ (triangles), and $.06e^{-.00011t}$ 
(squares). These exponentials are displayed by dashed lines in the figure. 
}
\end{figure}

The region of exponential fidelity decay varies dramatically with the 
location of the coherent state on the underlying classical phase space. This 
is displayed in Fig.~\ref{G2Exp2} where a wide range of exponential decay 
rates are found for different coherent states though they undergo 
equivalent evolution. In addition, the time of the transition period from 
Gaussian to exponential varies from state to state.

In conclusion, we have provided a numerical study of fidelity decay 
behavior for coherent states in a quantum system whose classical analog 
is quasi-integrable. We find that the initial fidelity decay 
behavior and rate will depend on the perturbation strength, 
Hilbert space dimension, and initial coherent state location. The quantum 
fidelity decay behavior generally corresponds to the classical fidelity 
decay explored in \cite{Ben2} and the prediction therein: quantum 
states tend more towards Gaussian decay due to the quantization of the 
phase space orbits. In addition, we show that the spectrum of the initial 
coherent state with respect to the system eigenstate extent contains 
information regarding the fidelity decay of that state. Finally, we find 
that after initial Gaussian decay behavior, there may be a second stage 
of exponential decay for strong perturbations. The rate and inception of the 
exponential decay depend on the perturbation strength and location of the 
coherent state. The existence of this second-stage decay behavior leads to the
counter-intuitive result that stronger perturbations may lead to higher 
fidelity, a phenomenon which may be important for quantum computation. 

The authors acknowledge support from the DARPA QuIST (MIPR 02 N699-00) 
program. Y.S.W. acknowledges the support of the National Research Council 
Research Associateship Program through the Naval Research Laboratory. 
Computations were performed at the ASC DoD Major Shared Resource Center.

\end{document}